\documentclass[10pt,aps,prd,amsmath,amssymb,superscriptaddress,onecolumn,nofootinbib,showpacs,preprintnumbers,amsfonts, notitlepage,hyperref={colorlinks,hyperindex, % pagebackref,
     bookmarks=true,bookmarksopen=true}]{revtex4-2}
\usepackage{slashed}
\usepackage{amsmath,amssymb}
\usepackage{bm}
\usepackage{multirow}
\usepackage{xspace}
\usepackage{colortbl}
\usepackage{ bbold }
\usepackage{comment}
\usepackage{upgreek}
\usepackage{orcidlink}
\usepackage{appendix}
\usepackage{tikz}
\usepackage{physics}
\usepackage{soul}
\usepackage{mathtools}

\usepackage{pifont}% http://ctan.org/pkg/pifont
\renewcommand{\arraystretch}{1.4}

\providecommand{\U}[1]{\protect\rule{.1in}{.1in}}
\newcommand{\gev}{\ensuremath{{\mathrm{\,Ge\kern -0.1em V}}}\xspace}
\newcommand{\gevsq}{\ensuremath{{\mathrm{\,Ge\kern -0.1em V}}^2}\xspace}
\newcommand{\mev}{\ensuremath{{\mathrm{\,Me\kern -0.1em V}}}\xspace}

\hypersetup{
colorlinks=true,
 }

\begin{document}

\title{Tensor meson couplings in AdS/QCD}

\newcommand{\kielce}{Institute of Physics, Jan Kochanowski University, ul. Uniwersytecka 7, 25-406, Kielce, Poland}

\newcommand{\bakub}{Institute of Physics, Ministry of Science and Education, H.Javid 33, Baku, AZ 1143,
Azerbaijan}

\newcommand{\bakua}{Institute for Physical Problems, Baku State University, Z.Khalilov 23, Baku, AZ-1148,
Azerbaijan}

\newcommand{\Khazar}{Center for Theoretical Physics, Khazar University, 41 Mehseti Street, Baku, AZ1096, Azerbaijan}

\newcommand{\rpi}{Institute of Radiation Problems, Ministry of Science and Education, B.Vahabzade 9, AZ1143, Baku, Azerbaijan}

\newcommand{\unec}{Department of Physics and Chemistry, “Composite Materials” Scientific Research Center, Azerbaijan State Economic University (UNEC), H. Aliyev 135, AZ1063, Baku, Azerbaijan}

\author{Shahin~Mamedov \orcidlink{0000-0002-0261-3914}\footnote{
\href{mailto:sh.mamedov62@gmail.com }{sh.mamedov62@gmail.com }}}
\affiliation{\bakua}\affiliation{\bakub}\affiliation{\Khazar}

\author{Zeynab~Hashimli 
\orcidlink{0009-0002-1230-8590}\footnote{
\href{mailto:z.hashimli21@gmail.com}{z.hashimli21@gmail.com}}}
\affiliation{\bakub}\affiliation{\Khazar}

\author{Shahriyar Jafarzade \orcidlink{0000-0002-4353-0760}\footnote{
\href{mailto:shahriyar.jzade@gmail.com}{shahriyar.jzade@gmail.com}}}
\affiliation{\kielce}\affiliation{\rpi}\affiliation{\unec}

\begin{abstract}
We study the hadronic and radiative couplings of the $f_2(1270)$ meson within the hard- and soft-wall models of AdS/QCD. The results for the tensor meson-nucleon-nucleon coupling ($g_{f_2NN}$) and tensor meson-photon-vector meson coupling ($g_{f_2\gamma\rho}$) are compared to the ones obtained by using the dispersion relations and amplitude methods, respectively. Qualitative agreement with different analyses implies the reliability of the holographic description of spin-2 meson. %Our results can be useful for the experiments devoted to the photoproduction of mesons, such as GlueX at Jefferson Lab. %The soft-wall model results are extended to the radially excited $f_2$ meson states too. 
\end{abstract}

\maketitle

\section{Introduction}
Quantum chromodynamics (QCD) describes the strong interaction within the Standard Model of particle physics. The binding of quarks into nucleons (protons and neutrons) and other hadrons, such as mesons composed of a quark-antiquark pair, is accomplished through strong interactions. One of the main challenges of QCD is the lack of a small dimensionless variable that would allow for the perturbative calculation of low-energy observables. Lattice QCD simulations and effective models are alternative approaches to studying low-energy QCD.  Phenomenological bottom-up approaches of holographic QCD are one of the effective approaches to low energy QCD which presume that the extra-dimensional fields propagate through a region of anti-de Sitter (AdS) space inspired by the AdS/CFT correspondence \cite{Maldacena:1997re}.

The holographic description of QCD observables was originally constructed to test the hadrons, including pseudoscalar, vector, and axial-vector mesons within the so-called hard-wall model (HW) of  AdS/QCD \cite{Erlich:2005qh}.  Three main experimental measurements, such as pion mass and decay constant and the mass of $\rho(770)$, were used to predict other quantities of the mesons within the model. 

Later on, the application of holographic QCD is extended to the spin-2 case within the HW model, in which tensor mesons were introduced similarly to gravitons  \cite{Katz:2005ir}. The equation of motion for the profile function of the $f_2(1270)$ meson was solved in this approach, and its decay into two pions and two photons was evaluated. Higher spin mesons, including $f_2(1270)$, were investigated within the SW model in \cite{Karch:2006pv}.

Nucleons within the HW model were studied in Ref. \cite{Hong:2006ta}. Further investigations include the coupling between mesons and nucleons, such as the nucleon coupling with the vector meson in Ref. \cite{Maru:2009ux}, with the axial-vector meson in Ref. \cite{Huseynova:2019bob} and with the pseudoscalar meson coupling in Ref.\cite{Kwee:2007nq}. Similar studies can be found in Refs. \cite{Ahn:2009px,Huseynova:2014pca, Kwee:2007dd, Wang:2015osq} and at finite temperature case in Refs.\cite{Mamedov:2021dpv,Mamedov:2021hkb}. %In this work, we will address the coupling between tensor mesons and nucleons. 

The resonance $f_2(1270)$ is the spin-$2$ meson with the positive parity (P) and charge conjugation (C) and the total spin number $J=2$.  The lightest spin-$2$ mesons $\{a_2(1320),K_2^\star(1430),f_2(1270),f_2(1525)\}$ make up a multiplet with quantum numbers  $J^{PC}=2^{++}$ and has a solid experimental \cite{Workman:2022ynf}, theoretical \cite{Dudek:2014qha} and phenomenological \cite{Giacosa:2005bw} foundation compatible with the quark model. Recent analyses of spin-2 mesons can be found e.g., in Refs \cite{Mathieu:2020zpm, Nys:2018vck, Bibrzycki:2013pja}. It has been studied within chiral perturbation theory in \cite{Chow:1997sg}. Especially, the mass and two-photon decay of $f_2(1270)$ estimates within holographic QCD fit the Particle Data Group (PDG) \cite{Workman:2022ynf} very well. Gravitational form factors of nucleons within AdS/QCD approach are computed in Refs.\cite{Abidin:2009hr,Abidin:2008ku, Abidin:2008hn} and extension to the finite temperature see Ref.\cite{Allahverdiyeva:2023fhn}. To what extent can holographic descriptions effectively portray various couplings? We address this question in this paper.  
%One of the remarkable features of this model is the ability to describe the mass and decay widths by using one single parameter. 

We first study the tensor meson nucleon coupling constant ($g_{f_2NN}$) within the hard- and soft-wall models of AdS/QCD and compare it to other known theoretical models, such as  Dispersion Relation (DR). Secondly, we estimate the tensor-vector-vector coupling ($g_{f_2\rho\rho}$) and the tensor-vector-photon coupling ($g_{f_2\gamma\rho}$) and compare the latter one to the results obtained from the CLAS experiment data \cite{CLAS:2009ngd} within amplitude methods. %Results for the couplings are extended to the radially excited tensor meson which can be useful to investigate the resonance of $f_2(1640)$ -a possible candidate for the radial excitation of $f_2(1270)$ \cite{Workman:2022ynf}.

The organization of the paper is as follows: Holographic formulations of mesons and nucleons are presented in Sec. \ref{meson} and Sec. \ref{nucleon}, respectively. Coupling constants between the tensor meson and nucleons are in Sec. \ref{results}. Sec. \ref{sec:tens-vect coupl} is devoted to tensor-vector-vector and tensor-vector-photon couplings. Our conclusion is in Sec. \ref{concluison}. An alternative derivation of the tensor meson profile function is discussed in App.\ref{App-A}, and a flavor-invariant form of the Lagrangian is presented in App.\ref{App-B}.

\section{Mesons in Ads/QCD}
\label{meson}
Background gravity for the AdS/QCD models is the 5D AdS spacetime, and the metric of it in Poincare coordinates has the following form:   
\begin{align}
    ds^2=\frac{1}{z^2}\Big(-dz^2+dx^\mu dx_\mu\Big) \quad (\mu=0,1,2,3),\label {eq:ads-spacetime}
\end{align}
where extra $z$ coordinate extends $0<z\leq z_m=323\, \text{GeV}^{-1}$ in the HW model and   $0<z<\infty$ in the SW model. The action of the model we are interested in has four components: free tensor field $h$, vector  field $v$, spinor field $N$, and their interactions:
\begin{align}    S=S_h+S_{v}+S_{N}+S_{int}.
\end{align}
In the next sections, we shall present the profile functions of these fields and perform calculations of the couplings within both hard- and soft-wall models.
\subsection{Tensor and vector mesons in the HW model}
 Tensor mesons in AdS/QCD were introduced in an analogy with the graviton field in the bulk of the AdS spacetime \cite{Katz:2005ir}, and here we present this solution briefly. Let us consider the 5D extension of the massive spin-2 field $h_{MN}(x,z)$ $(M, N=0,1,2,3,5)$. The gauge condition on fifth components of $h_{MN}$ implies 
 \begin{align}
   h_{MN}= 
\begin{dcases}
    0,& \text{if } $M$ \,\text{or}\, $N=5$\\
    h_{\mu\nu},              & \text{otherwise}
\end{dcases}
 \end{align}
The linear perturbation of gravity around the AdS background is given by the $h_{\mu\nu}(x,z)$ field:
	\begin{align}
	    ds^2=\dfrac{1}{z^2}(\eta_{\mu\nu}+h_{\mu\nu})dx^ \mu dx^\nu -\dfrac{1}{z^2}dz^2\,,
	\end{align}
leading to the following action:
	\begin{align}
		S_h^{HW}=-\dfrac{1}{4}\int d^{5}x \dfrac{1}{z^{3}}\Big[(\partial_z h_{\mu\nu})(\partial_z h^{\mu\nu})+h_{\mu\nu}\Box h^{\mu\nu}+\text{higher order terms}\Big].
	\end{align}
 The UV boundary value of $h_{\mu\nu}$ corresponds to the wave function of the graviton. The first mode in the Kaluza-Klein (KK) decomposition of $h_{\mu\nu}$ will be the lightest tensor meson listed in the introduction, i.e., the $f_{2}(1270)$ meson.  %\textcolor{red}{Following Ref. \cite{Katz:2005ir} the kinetic term of the Lagrangian for the $h_{MN}$ tensor field is written as an extension to the five-dimensional spacetime (\ref{1}) of the four-dimensional Fierz-Pauli Lagrangian \cite{Fierz:1939ix} 
%\begin{align}
%L_{kin}^{(f)}=\dfrac{1}{2}h_{\mu\nu} \Box\, h_{\mu\nu}+h_{\mu\alpha,\alpha}^{2}-h_{\mu\nu,\mu}h_{,\mu}+\dfrac{1}{2}h_{,\mu}^{2}+\dfrac{1}{2}m_{f}^{2}(h_{\mu\nu}^{2}-h^{2}), \label{5}
%\end{align}
%where $h=h_{\mu\mu}$ .}
%But since we are not demanding a fully consistent theory, we could just as well have introduced a different spin-2 field.
QCD predicts the existence of multiple isospin singlets spin-2 particles \footnote{ See for instance Ref.\cite{Vereijken:2023jor} for the recent detailed discussion about spin-2 isoscalars.}, such as glueballs. Therefore, a comprehensive formulation of the spin-2 particles should contain various bulk spin-2 fields. Each spin-2 particle appears as a KK mode $\Big(h_{\mu\nu}(x,z)=\sum_n h_{\mu\nu}^{(n)}(x)h^{HW}_{n}(z)\Big)$ where the profile function $h^{HW}_{n}(z)$ satisfy the following equation of motion:
\begin{align}
	\partial^2_z h^{HW}_{n}(z)-\dfrac{3}{z}\partial_zh^{HW}_{n}(z)+(m_{n}^{h})^{2}h^{HW}_{n}(z)=0.
\end{align}

The generic solution to this equation is expressed in terms of Bessel functions of the first kind $J_{2}$ and the second kind  $Y_{2} $ \cite{Katz:2005ir}:
\begin{align}
	h_{n}^{HW}(z)=N_{n}z^{2}[J_{2}(m_{n}^{h}z)+\beta_{n}Y_{2}(m_{n}^{h}z)]. 
\end{align}
  UV boundary condition $ h^{HW}(0)=0 $ implies $ \beta_{n}=0 $ and the IR boundary condition $h^{HW\prime}(z_m)=0$ quantize the tensor meson masses according to $ J_1 (m_{n}^{h}z_{m})=0 $. Thus the profile function for the $ f_2(1270) $ meson is
\begin{align}
	h^{HW}(z)=3.51\frac{z^{2}}{z_{m}}J_{2}(3.83\frac{z}{z_{m}}) 
 \text{ with the normalization: } \int_0^{z_m} dz\Big(h^{HW}_n(z)\Big)^2 / z^3=1 . 
\end{align}
The $ f_2(1270)$ mass is predicted by the IR boundary condition of this solution and equals to $ 3.83/z_{m}=1236$ MeV, which is only $ 3 \% $  off of the observed mass in \cite{Workman:2022ynf}. However, predictions for the two pion decays are underestimated compared to the PDG \cite{Workman:2022ynf}.

Vector mesons in the HW AdS/QCD are introduced using a 5D vector field composed of the left and right gauge fields. The action for the $V_M$ vector field is
\begin{align}\label{eq:vect-action}
    S_v^{HW}=-\frac{1}{4g_5^2}\int d^5x\sqrt{g} V^{MN}V_{MN}=-\dfrac{1}{4g_5^2}\int d^5x (V^2_{\mu\nu}-2V^2_{\mu5})/z\,,
\end{align}
where $V_{\mu\nu}:= \partial_\mu V_\nu-\partial_\nu V_\mu-i[V_{\mu},V_{\nu}]$, $V_{\mu}:=V_{\mu}^at^a$ and the coupling $g_5=\sqrt{4\pi^2N_f/N_c}$ was obtained in Ref.\cite{Erlich:2005qh}. The equation of motion for KK modes $\Big(V_{\mu}(x,z)=\sum_n V_{\mu}^{(n)}(x)v^{HW}_n(z)\Big) $ of the vector field in the HW model has the following form 
\begin{align}
    \partial^2_zv^{HW}_n(z)- \dfrac{1}{z}\partial_z v^{HW}_n(z)+ (m^v_n)^2v^{HW}_n(z)=0\,,
\end{align}
and solution to this equation is also expressed in terms of Bessel functions:
\begin{align}
    v^{HW}_n(z)=N_nz\Big[J_1(m^v_nz)+\beta_n Y_1(m^v_nz)\Big]. 
\end{align}
One can use boundary conditions $v^{HW\prime}(z_m)=v^{HW}(0)=0$, which sets $\beta_n=0$ and quantize the masses of the vector mesons according to the IR condition $J_0(m^V_nz_m)=0$. Normalization is set by
\begin{align}
    \int_0^{z_m}  \dfrac{  dz }{z}\Big(v^{HW}_n (z)\Big)^2=1\,, \label{eq:norm-vec}
\end{align}
and the ground state profile function of $\rho$ meson reads
\begin{align}
    v^{HW}(z)\equiv v^{HW}_0(z)=2.72\dfrac{z}{z_m}J_1(2.4\dfrac{z}{z_m})\,,\label{eq:prof-vec}
\end{align}
where $z_m=(323 \text{MeV})^{-1}$ is fixed with experimental mass of $\rho(770)$ meson.

\subsection{Tensor and vector meson profile functions in the SW model}
 In the SW model, the spin-$S$ mesons are described by the rank-$S$ tensor field $\phi^{\mu_1...\mu_S}(x,z)$ in the bulk of the 5D AdS space and the quadratic part of the gauge and coordinate invariant action for this field is written in the form \cite{Karch:2006pv}:
 \begin{align}
     S_{v/h}^{SW}=\dfrac{1}{2}\int d^5x\, \exp{-\Big(\Phi(z)-A(z)(2S-1)\Big)} \partial_N \phi_{\mu_1...\mu_S}\partial^N \phi^{\mu_1...\mu_S} .
 \end{align}
Here $A$ denotes $A(z)=-\log kz$ and the dilaton field $\Phi(z)=k^2z^2$ is introduced to regulate the integral over the extra dimension $z$ and to give a meson spectra agreeing with Regge trajectory. The 5D equation of motion for the KK modes $\phi_n $ of the transverse traceless part of the $\phi_{\mu_1...\mu_S}$ field  $\Big(\phi_{\mu_1...\mu_S}(x,z)=\sum_n \phi_n(x,z) \phi^{(n)}_{\mu_1...\mu_S}(x)\Big)$  can be easily derived from this action by considering  $\Box \phi^{(n)}_{\mu_1...\mu_S}(x)=m^2_n \phi^{(n)}_{\mu_1...\mu_S}(x)$:
\begin{align}
    \partial_z(e^{-(\Phi(z)-A(z)(2S-1))}\partial_z\phi_n(z))+m^2_ne^{-(\Phi(z)-A(z)(2S-1))}\phi_n(z)=0.
    \end{align}
 By making  substitution $\phi_n(z)= e^{{-(\Phi(z)-A(z)(2S-1))}/2}\Tilde{\phi}_n(z)$ in the equation it gives the Schrödinger-like equation form for $\Tilde{\phi}$
\begin{align}
    -\Tilde{\phi}''_n(z)+\Big[k^4z^2+2(S-1)k^2+\dfrac{4S^2-1}{4z^2}\Big]\Tilde{\phi}_n(z)=E\Tilde{\phi}_n(z).    \label{eq: eom-soft-meson}
\end{align}
%where the potential $V(z)$ is:
%\begin{align}
%    V(z)=k^4z^2+2(S-1)k^2+\dfrac{4S^2-1}{4z^2}.
 %   \label{6}
%\end{align}
Solving the Eq. (\ref{eq: eom-soft-meson}) will give the eigenfunctions, which are so-called the profile functions for the spin-$S$ mesons, and the corresponding eigenvalues are
\begin{align}
    E=4n-2S+2\,,  
\end{align}
with the mass spectrum
\begin{align}
    m_{n,S}^2=4k^2(n+S).
\end{align}
\begin{enumerate}
    \item 
 For the vector meson ($S=1$) the equation of motion (\ref{eq: eom-soft-meson}) gets a form:
\begin{align}
    -\Tilde{\phi}''_n(z)+\Big[k^4z^2+\dfrac{3}{4z^2}\Big]\Tilde{\phi}_n(z)=m_n^2 \Tilde{\phi}_n(z).
\end{align}
By solving this equation with the substitution $v_n^{SW}(z)=\sqrt{kz}e^{k^2z^2/2}\Tilde{\phi}_n(z)$  the profile function in terms of Laguerre polynomials ($L_n^1(x)$) reads:
\begin{align}    v^{SW}_n(z)=\sqrt{\dfrac{2n!}{(n+1)!}} k^2z^2 L_n^1(k^2z^2).
\end{align}
Note that, parameter $k$ is fixed according to the mass of $\rho(770)$ which leads $k=0.388\,\text{GeV}$.
\item For the tensor meson ($S=2$)  the equation of motion (\ref{eq: eom-soft-meson}) becomes
\begin{align}\label{eq:eom:sw-ten}
    -\Tilde{\phi}''_n(z)+\Big[2k^2+k^4z^2+\dfrac{15}{4z^2}\Big]\Tilde{\phi}_n(z)=m_n^2\Tilde{\phi}_n(z)\,,
\end{align}
and the substitution $h^{SW}_n(z)=\sqrt{(kz)^3}e^{k^2z^2/2}\Tilde{\phi}_n(z)$ leads to the following profile function:
\begin{align}\label{eq:sw-ten-prof-fun}
   h_n^{SW}(z)=(kz)^4\sqrt{\frac{2n!}{(n+2)!}}L_n^2(k^2z^2),
   \end{align}
which has the normalization:
\begin{align}    \int_{0}^{\infty}dz e^{-k^2z^2}\,\Big(h_n^{SW}(z)\Big)^2/z^3=1.
\end{align}
In the appendix \ref{App-A}, we show that the solution for the tensor meson profile function in \eqref{eq:sw-ten-prof-fun} is the same as the one derived in \cite{Abidin:2009hr} if one approximates the spin-2 tensor meson as a graviton.

\end{enumerate}
\section{Nucleons in Ads/QCD }\label{nucleon}

\subsection{Nucleon profile function in the HW model}
Two bulk fermion fields, $N_1^{HW}$ and $N_2^{HW}$ are required in the AdS space-time to describe the independent left and right chiral components of the boundary nucleon \cite{Hong:2006ta}. The Lagrangian for these spinors has different signs for the "5-dimensional mass" $m_5$, which is related to the conformal dimension of the spinor: $m_5=\dfrac{5}{2}$ for $N_1^{HW}$ and $m_5=-\dfrac{5}{2}$ for $N_2^{HW}$. In light of the normalizability and chirality analyses, $N_1^{HW}$ describes the left-handed component of the boundary nucleon, while $N_2^{HW}$ describes the right-handed one. The IR boundary condition eliminates the remaining components of the spinors and gives the mass spectrum of the nucleon $m_n$. The bulk action for the Dirac spinors $ N_1^{HW} $ and $ N_2^{HW} $ is given by \cite{Hong:2006ta}:
	\begin{align}\nonumber
	     S_{N}^{HW}=\int d^5x\sqrt{g}\Bigg[\dfrac{i}{2}\overline{N}^{HW}_1e^M_A\Gamma^A\nabla_MN^{HW}_1-\dfrac{i}{2}\Big(\nabla^\dagger_M\overline{N}^{HW}_1\Big)\,e^M_A\Gamma^AN^{HW}_1&-m_5\overline{N}^{HW}_1N^{HW}_1\\
      &+\Big(N_1^{HW}\rightarrow N_2^{HW}, m_5\rightarrow -m_5\Big)\Bigg], \label{eq: action-nucleon-hard}
	\end{align}
	where $e^A_M=\dfrac{1}{z}\eta ^A_M$ is the vielbein and the Dirac matrices $ \Gamma^A=\Big(\gamma^{\mu},-i\gamma^{5}\Big)$ satisfy the Clifford algebra $ \Big\{\Gamma^A,\Gamma^B\Big\}=2\eta^{AB} $. 
	The Lorentz and gauge covariant derivative is:
	\begin{align}
	    \nabla_M=\partial_M+\dfrac{i}{4}\omega^{AB}_M\Gamma_{AB}, \label{eq: covariant-derivative}
	\end{align}
	where the Lorentz generator $\Gamma^{AB}$ is defined as $\Gamma^{AB}:=\dfrac{1}{2i}\Big[\Gamma^A,\Gamma^B\Big]$  and the non-vanishing components of the spin connection $\omega^{AB}_M$ equal to $\omega^{5A}_M=-\omega^{A5}_M=\dfrac{1}{z}\delta^A_M$.
	Extremizing the action (\ref{eq: action-nucleon-hard}) with respect to $\overline{N}_1^{HW}$, we get 5D Dirac equation for $N_1^{HW}$: 
	\begin{align}
	    (ie^M_A \Gamma^A \nabla_M-m_5)N_1^{HW}=0\,,
     \label{eq: dirac-nucleon-hard}
	    	\end{align}
with the boundary condition
 \begin{align}
     \Big[\delta\overline{N}^{HW}_1e^5_A\Gamma^AN^{HW}_1\Big]^{z_m}_{\epsilon}=0. 
 \end{align}
Similarly, for the field of $N_2^{HW}$.
	Since the AdS/CFT correspondence relates the classical solution to the bulk equation of motion with the corresponding boundary operator, it is useful to decompose the bulk spinor into left and right components, similar to the 4D chirality projectors, when evaluated at the UV brane:
	\begin{align}
	    N_{1}^{HW}(x,z)=N_{1L}^{HW}(x,z)+N_{1R}^{HW}(x,z) \,, \qquad N_{2}^{HW}(x,z)=N_{2L}^{HW}(x,z)+N_{2R}^{HW}(x,z) \label{eq: psi-left-right},
     \end{align}
where  $N_{1,2L/R}^{HW}=\dfrac{1\mp i\Gamma^5}{2}N_{1,2}^{HW}$ are chiral spinors. In momentum space, they are written  in terms of the left (right) 4D spinors $u_{L,R}(p)$ satisfying the 4D Dirac equation 
\begin{align}
     N_1^{HW}(p,z)=F^{HW}_{1L}(p,z)  {u}_L(p)+F_{1R}^{HW}(p,z){u}_R(p)\,,\qquad
    N_2^{HW}(p,z)=F_{2L}^{HW} (p,z) {u}_L(p)+F_{2R}^{HW} (p,z) {u}_R(p).
\end{align}
We perform KK decomposition for the fermion fields as follows:
\begin{align}
    N_{1,2L/R}^{HW}(p,z)=\sum_n N^n_{1,2L/R}(p) F^{n\,HW}_{1,2L/R}(p,z) \label{eq: psi-KK}.
\end{align}
%where $N^n_{1,2L/R}$ chiral spinors \st{solve 4D Dirac equation} . 
Mode functions for ground state nucleons $F^{HW}_{L,R}(z)\equiv F^{0\,HW}_{L,R}(z) $ are the solutions of the Dirac equation (\ref{eq: dirac-nucleon-hard}) \cite{Hong:2006ta}
\begin{align}
	F_{1L}^{HW}(p,z)=c_{1}z^{5/2}J_{2}(m_{n}z)\,,\qquad  F_{1R}^{HW}(p,z)=c_{1}z^{5/2}J_{3}(m_{n}z) ; 
\end{align}
\begin{align}
	F_{2L}^{HW}(p,z)=-c_{2}z^{5/2}J_{3}(m_{n}z)\,,\qquad F_{2R}^{HW}(p,z)=c_{2}z^{5/2}J_{2}(m_{n}z), 
\end{align}
with the normalization constants \cite{Maru:2009ux} :
\begin{align}
	|c_{1,2}|=\dfrac{\sqrt{2}}{z_{m}J_{2}(m_{n}z)}. 
\end{align}

\subsection{Nucleon profile function in the SW model}
In the SW model with the negative dilaton field $\Phi$ the action for the 5D $N_{1}$ spinor fields with $m_5=5/2$ (similarly for $N_{2}$ with $m_5=-5/2$  )  is written as \cite{Gutsche:2011vb}:
\begin{align}\nonumber
    S_{N_1}^{SW}=\int d^4xdz \sqrt{g}e^{-\Phi(z)} \Big [\dfrac{i}{2}\overline{N}_1^{SW}(x,z)e^M_A\Gamma^A \nabla_M N_1^{SW}(x,z)-\dfrac{i}{2}\Big(\nabla_M N_1^{SW}(x,z)\Big)^\dagger\Gamma^0 e^M_A\Gamma^A  N_1^{SW}(x,z) 
    \\
    -\overline{N}_1^{SW}(x,z)\Big(m_5+\Phi(z)\Big)N_1^{SW}(x,z)\Big]\,.
\end{align}
For removing the dilaton field from the overall exponential, the fermionic fields are rescaled as $N^{SW}_{1,2}(x,z) \rightarrow e^{k^2z^2/2}N^{SW}_{1,2}(x,z)$.  We decompose the fermion field in left- and right-chirality components (\ref{eq: psi-left-right}) and KK modes \footnote{Since indices 1 and 2 follow the same analogical structure as the HW model, we shall omit them in order to avoid repetition.}. To obtain the equation of motion, we use the following substitution: 
\begin{align}
    F^{n\,{SW}}_{L/R}(z)=z^{2}f^{n\,{SW}}_{L/R}(z). 
\end{align}
The equation of motion for fermions can be represented as a Schrödinger-type equation
\begin{align} 
    \Big[-\partial^2_z+k^4z^2+2k^2\Big(m\mp \dfrac{1}{2}\Big)+\dfrac{m(m\pm 1)}{z^2} \Big]f^{n\,{SW}}_{L/R}(z)=m^2_n f^{n\,{SW}}_{L/R}(z). \label{eq:eom-nucleon-soft}
\end{align}
The solution of the Eq. (\ref{eq:eom-nucleon-soft}) gives the mode functions for nucleons in the SW model:
\begin{align} 
    f^{n\,{SW}}_L(z)=\sqrt{\dfrac{2\Gamma(n+1)}{\Gamma(n+3)}} k^{3}z^{5/2}e^{-k^2z^2/2}L^{2}_n(k^2z^2), \,\qquad
    f^{n\,{SW}}_R(z)=\sqrt{\dfrac{2\Gamma(n+1)}{\Gamma(n+2)}} k^{2}z^{3/2}e^{-k^2z^2/2}L^{1}_n(k^2z^2). 
\end{align}
Undoing the change of variables, we get the original mode functions
\begin{align}
     F^{n\,{SW}}_L(z)=\sqrt{\dfrac{2\Gamma(n+1)}{\Gamma(n+3)}} k^{3}z^{9/2}L^{2}_n(k^2z^2), \,\qquad
    F^{n\,{SW}}_R(z)=\sqrt{\dfrac{2\Gamma(n+1)}{\Gamma(n+2)}} k^{2}z^{7/2}L^{1}_n(k^2z^2)\,.
\end{align}
The normalization is set by 
\begin{align}    \int_{0}^{\infty}\dfrac{e^{k^2z^2}}{z^4} F^{n\,{SW}}_{L/R}(z)F^{m\,{SW}}_{L/R}(z)=\delta^{mn}\,,
\end{align}
and the mass spectrum of the nucleon will be given by
\begin{align}
    m^2_n=4k^2(n+2). 
\end{align}
Note that we will consider the interaction only for the ground states $n=0$ and will use the short-hand notation for the SW profile functions as 
\begin{align}
   F_{L/R}^{\,SW}(z)\equiv  F_{L/R}^{0\,SW}(z)\,.
\end{align}

\section{Tensor-Nucleon couplings}\label{results}

In order to derive the tensor meson-nucleon coupling constant $g_{f_{2}NN}$ within the holographic QCD, we have to write down the interaction Lagrangian $L^{5D}_{f_{2}NN}$ for the corresponding fields in the bulk of the 5D AdS space. The interaction Lagrangian  between the tensor meson and nucleons in 4D theory is given in  Ref.  \cite{Yu:2011zu}:
	\begin{align}
	L_{f_{2}NN}=-2i\frac{g^{(1)}_{f_{2}NN}}{m_n}\overline{N}(x)(\gamma^\mu \overleftrightarrow{\partial}^\nu+\gamma^\nu \overleftrightarrow{\partial}^\mu)N(x) h_{\mu \nu}(x)+4\frac{g^{(2)}_{f_{2}NN}}{m_n^2} \partial^\mu \overline{N}(x) \partial^\nu N(x) h_{\mu \nu}(x)\,,\label{eq: Lagranjian-fNN}
	\end{align}
where $\overleftrightarrow{\partial}=\overleftarrow{\partial}-\overrightarrow{\partial}$ and $M_N$ denote the mass of nucleon. It is reasonable to write 4D action in momentum space using Fourier transforms of the nucleon $N(x)$ and tensor meson $h_{\mu\nu}(x)$ fields:
%	\begin{align}
%	     S^{4D}=-2i\frac{g^{(1)}_{f_{2}NN}}{M} \int d^4 x \overline{N}(\gamma^\mu \overleftrightarrow{\partial^\nu}+\gamma^\nu \overleftrightarrow{\partial^\mu})N h_{\mu \nu}+4\frac{g^{(2)}_{f_{2}NN}}{M^2} \int d^4 x \partial^\mu \overline{N} \partial^\nu N h_{\mu \nu}
%	\end{align}
	\begin{align}
	    \overline{N}(x)=\int d^4p ' \overline{N}(p') e^{ip' x}\,,\qquad
	    N(x)=\int d^4pN(p) e^{-ipx}\,,\qquad 
	    h_{\mu \nu}(x)=\int d^4qh_{\mu \nu}(q)e^{-iqx}. 
	\end{align}
The action for the interaction of these fields in momentum space has two terms:
\begin{align}    S^{4D}_{\text{int}}=4\frac{g^{(1)}_{f_{2}NN}}{m_n}\int d^4p' d^4p(p'+p)^\nu\overline{N}(p')\gamma^\mu N(p)h_{\mu\nu}(q)+4\frac{g^{(2)}_{f_{2}NN}}{m_n^2}\int d^4p' d^4p p'^\mu p^\nu \overline{N}(p')N(p)h_{\mu\nu}(q), \label{eq: action-fNN-4D}
\end{align}
which are proportional to $(p'+p)^{\nu}\gamma^{\mu}$ and $p'^{\mu}p^{\nu}$ in the tensor current of the nucleon. In 5D space-time (\ref{eq:ads-spacetime}), the interaction Lagrangian between the bulk tensor $h_{MN}$ and fermionic fields $N$ is the extension of the Lagrangian (\ref{eq: Lagranjian-fNN}). To this end, we should replace partial derivatives with the covariant derivatives (\ref{eq: covariant-derivative}) of the spinor field and construct hermitian Lagrangian. Thus, we can write the 5D action 
\begin{align}
S_{\text{int}}^{5D}=\int d^5x\sqrt{g}\, L^{5D}_{f_2NN}\,,
\end{align}
based on the following interaction Lagrangian:
\begin{align}
L_{f_2NN}^{5D}&=2i\overline{N}(x,z)\Big(\Gamma^M \overleftrightarrow{\nabla}^N+\Gamma^N \overleftrightarrow{\nabla}^M\Big)N(x,z) h_{MN}(x,z) \nonumber
    \\
    &+4\Bigg(N(x,z) ^\dagger\Gamma^0\nabla^M\nabla^NN(x,z)+\Big(N(x,z) ^\dagger\Gamma^0\nabla^M\nabla^NN(x,z)\Big)^\dagger\Bigg) h_{MN}(x,z)\,, 
\end{align}
where $\overleftrightarrow{\nabla}=\overleftarrow\nabla-\overrightarrow\nabla$. 
Note that the second term with two derivatives is written in such a way as to ensure the hermiticity of the Lagrangian. One can show that it is a correct 5D generalization of the corresponding 4D term, and it looks like to the 4D Lagrangian upon replacing $(\mu,\nu\rightarrow M,N)$
\begin{align}
L^{5D}_{f_2NN}= 2i\overline{N}(x,z)\Big(\Gamma^M \overleftrightarrow{\partial}^N+\Gamma^N\overleftrightarrow{\partial}^M\Big)N(x,z) h_{MN}(x,z)
    + 4 \partial^M \overline{N}(x,z) \partial^N N(x,z) h_{M N}(x,z)\,.
\end{align}

In momentum space, 5D action in the HW model reads:
\begin{align}    S^{HW}_{\text{int}}=2\int_0^{z_m}  \dfrac{dz}{z^4} \Big\{\Big(F^{HW}_{1L}(z)\Big)^2+\Big(F^{HW}_{1R}(z)\Big)^2\Big\}h^{HW}(z)\int d^4p' d^4p(p^{\nu\prime}+p^\nu)\overline{u}(p')\gamma^\mu u(p)h_{\mu\nu}(q) \nonumber
  \\
 +2\int_{0}^{z_m}  \dfrac{dz}{z^5}\Big\{\Big(F^{HW}_{1L}(z)\Big)^2-\Big(F^{HW}_{1R}(z)\Big)^2\Big\}h^{HW}(z)\int d^4p' d^4p p^{\mu\prime} p^\nu \overline{u}(p')u(p)h_{\mu\nu}(q).\label{eq: action-fNN-5D-hard}
\end{align}

Comparing equations \eqref{eq: action-fNN-4D} and \eqref{eq: action-fNN-5D-hard}, we find the coupling constants $g^{(1)}_{f_2NN}$ and $g^{(2)}_{f_2NN}$ for the interaction between the ground state nucleon and the tensor meson. In the HW model, we find the following integral representations for these couplings:
\begin{align}
    g^{(1)\,HW}_{f_2NN}=\dfrac{m_n}{2}\int_{0}^{z_m}  \dfrac{dz}{z^4}\Big\{\Big(F^{HW}_{1L}(z)\Big)^2+\Big(F^{HW}_{1R}(z)\Big)^2\Big\}h^{HW}(z),\label{eq:g1_fNN-hard}
    \\
    g^{(2)\,HW}_{f_2NN}=\dfrac{m^2_n}{2}\int_{0}^{z_m} \dfrac{dz}{z^5}\Big\{\Big(F^{HW}_{1L}(z)\Big)^2-\Big(F^{HW}_{1R}(z)\Big)^2\Big\}h^{HW}(z).\label{eq:g2_fNN-hard}
\end{align}

Action in the SW model gets the following form:
\begin{align}   S^{SW}_{\text{int}}=2\int_{0}^{\infty} dz  \dfrac{e^{-k^2z^2}}{z^4} \Big\{\Big(F^{SW}_{L}(z)\Big)^2+\Big(F^{SW}_{R}(z)\Big)^2\Big\}h^{SW}(z)\int d^4p' d^4p(p+p^{\prime})^\nu\overline{u}(p')\gamma^\mu u(p)h_{\mu\nu}(q) \nonumber
   \\
   +2\int_{0}^{\infty} dz \dfrac{e^{-k^2z^2}}{z^5}\Big\{\Big(F^{SW}_{L}(z)\Big)^2-\Big(F^{SW}_{R}(z)\Big)^2\Big\}h^{SW}(z)\int d^4p' d^4p p^{\prime\mu} p^\nu \overline{u}(p' )u(p)h_{\mu\nu}(q). 
\end{align}

The couplings are as follows:
\begin{align}
    g^{(1)\, SW}_{f_2NN}=\dfrac{m_n}{2}\int_{0}^{\infty} dz \dfrac{e^{-k^2z^2}}{z^4} \Big\{\Big(F^{SW}_{L}(z)\Big)^2+\Big(F^{SW}_{R}(z)\Big)^2\Big\}h^{SW}(z), \label{eq:g1_fNN-soft}
    \\
    g^{(2)\, SW}_{f_2NN}=\dfrac{m^2_n}{2}\int_{0}^{\infty} dz\dfrac{e^{-k^2z^2}}{z^5}\Big\{\Big(F^{SW}_{L}(z)\Big)^2-\Big(F^{SW}_{R}(z)\Big)^2\Big\}h^{SW}(z). \label{eq: g2_fNN-soft}
\end{align}
In Table \ref{tab:nucleon-couplings}, we present the results of our numerical calculations compared to the analyses performed in various works.
%In Tab.\ref{tab:nucleon-couplings}, the values of these couplings, which have been found here and within the tensor meson dominance (TMD) and dispersion relations (DR) approaches have been given. As is seen from the table the AdS/QCD models give results close to the ones obtained within the TMD and DR approach calculations.

\begin{table}[h]
\begin{tabular}{|c|c|c|c|c|c|c|c|}
\hline\renewcommand{\arraystretch}{2.}
\centering
Couplings & Hard Wall & Soft Wall & TMD\cite{Renner:1971sj} & DR \cite{Engels:1970htu}  &DR \cite{Goldberg:1968zza}  &  DR \cite{Oh:2003aw} & DR \cite{Borie:1976nv} \\ \hline
$|g^{(1)}_{f_2NN}|$      &  $1.06$         &    8.46       & $2.18\pm 0.12$ & $6.44\pm 0.60$  & $3.75$ & $5.16$ & $5.26\pm 2.15$  \\ \hline
$|g^{(2)}_{f_2NN}|$      &    $0.23$       &  0.42       & $\approx 0$   & $0.39\pm 1.10$ & $-$  &  $\approx0$ & $3.15\pm 4.73 $  \\ \hline
\end{tabular}
\caption{Predictions for the couplings compared to the Refs.\cite{Renner:1971sj,Engels:1970htu,Goldberg:1968zza,Oh:2003aw,Borie:1976nv}}
\label{tab:nucleon-couplings}
\end{table}
\begin{figure}[h]
        \centering       \includegraphics[scale=0.55]{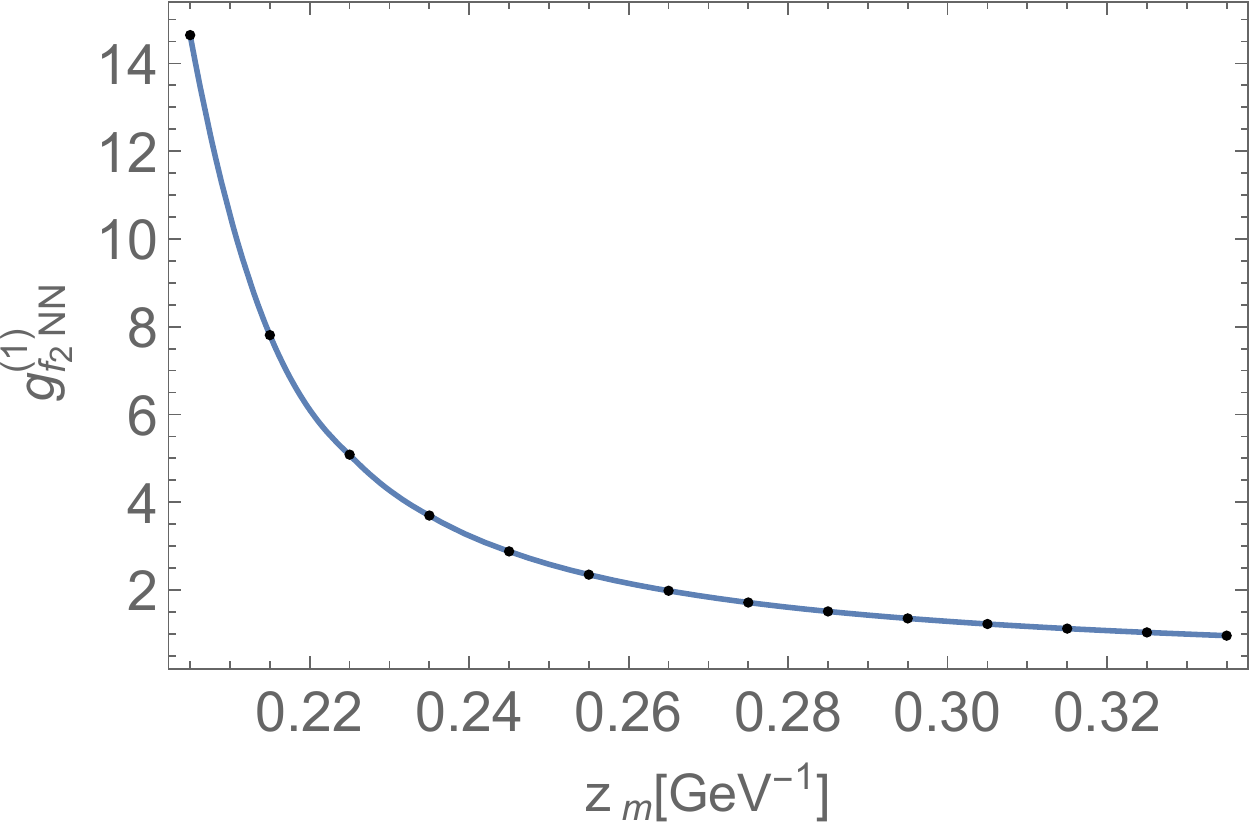}~~~ \includegraphics[scale=0.45]{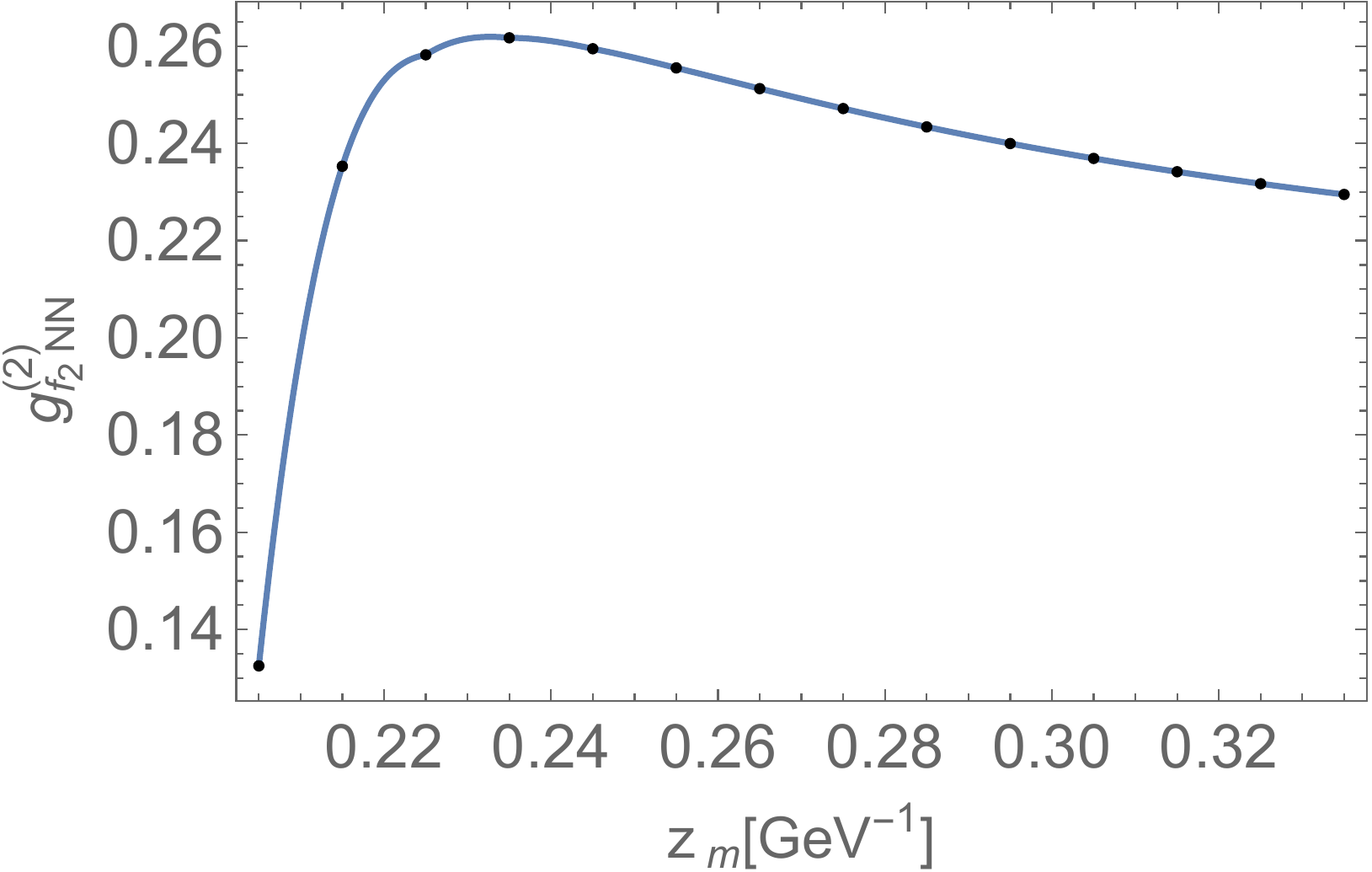} \\
        \caption{Dependence of the couplings on the cutoff value $z_m$ within the HW model}
        \label{fig}
   \end{figure}
It is interesting to know the sensitivity of the couplings to the value of the cutoff parameter $z_m$. To this end, we plot in Fig.\ref{fig} the dependencies of the $g^{(i)}_{f_2NN}$ constants on $z_m$. These plots are obtained by interpolating the points (black dots) corresponding to the different cutoff values. It is seen from Fig.\ref{fig}, in the domain approximately $0<z_m<0.24$ $\text{GeV}^{-1} $, the result for the couplings is susceptible to the cutoff values $z_m$. In the domain $0.24\, \text{GeV}^{-1}<z_m<\infty$  the couplings weakly depend on this parameter. In the case of the $z_m=0.322$ $\text{GeV}^{-1}$ in which the masses of mesons are fixed to the experiment, we observe some deviation from other results. While in the case of  $z_m=0.205$ $\text{GeV}^{-1} $ in which nucleons are well described, we get $g^{(1)}_{f_2NN}\approx 15$. Therefore, all results obtained by different groups can be described within the HW model in the domain of $z_m\in [0.205,0.322]\, \text{GeV}^{-1}$.   Note that some underestimate for tensor-nucleon coupling can be understood in the context of Tensor Meson Dominance (TMD) since the AdS/QCD prediction for $g^{\text{HW}}_{f_2\pi\pi}=0.009$ is two times less than the $SU(3)$ effective model prediction $g^{SU(3)}_{f_2\pi\pi}=0.019$ \cite{Renner:1971sj}. If we set $g^{(2)}_{f_2NN}=0$ in the following relation (see derivation in the appendix of Ref.\cite{Yu:2011zu}), 
\begin{align}
    \frac{2}{m_n}\Big( g^{(1)}_{f_2NN}+g^{(2)}_{f_2NN} \Big)=\frac{g_{f_2\pi\pi}}{m_{f_2}}\label{53}
\end{align}
we get $g^{(1)}_{f_2NN}=1.06$, which is approximately two times less than in the Ref. \cite{Renner:1971sj}.
%\section{tensor meson-vector meson coupling within ads/qcd}
%\begin{align}
 %   L_{f_2\pi\pi}=\dfrac{1}{2}g_{f_2\pi\pi}h_{\mu\nu}(\dfrac{1}{2}\eta_{\mu\nu}(\partial_\alpha \pi)^2-(\partial_\mu \pi)(\partial_\nu \pi))
%\end{align}
%\begin{align}
 %   S=\int d^4x dz \Big\{ -\dfrac{1}{4z}(F^L_{MN})^2-\dfrac{1}{4z}(F^R_{MN})^2+\dfrac{v(z)^2}{2z^3}(A_M)^2 \Big\}
%\end{align}
%\begin{align}
 %   g_{f\pi\pi}^{\text{HW}}=\frac{4\pi}{\sqrt{5}}\int_0^{z_m} dz h(z)\Big\{\frac{1}{z}\pi(z)^2+\frac{z^3}{2v(z)^2}\Big(\partial_z\frac{\pi(z)}{z}\Big)^2\Big\}=9.05\cdot 10^{-3} \,\text{MeV}^{-1}
%\end{align}
%\begin{align}
 %   g_{f\pi\pi}^{\text{SW}}=\int_0^{\infty} dz e^{-k^2z^2}h(z)\Big\{\frac{1}{z}\pi(z)^2+\frac{z^3}{2v(z)^2}\Big(\partial_z\frac{\pi(z)}{z}\Big)^2\Big\}=13\cdot 10^{-3} \,\text{MeV}^{-1}
%\end{align}

\section{Tensor meson decay constants $g_{f_2\rho \rho}$ and $g_{f_2\gamma \rho }$}  \label{sec:tens-vect coupl}
The tensor meson-vector meson interaction, which includes $f_2$ decays to the vector particles $\rho$ and $\gamma$ ( $f_2\rightarrow\rho\rho$, $f_2\rightarrow \gamma\rho$, and $f_2 \rightarrow \gamma\gamma$) lies in the bulk action for the vector field in the background of Eq. \eqref{eq:vect-action}.
%\begin{align}
 %   S=-\dfrac{1}{4g^2_5}\int d^5 x \sqrt{g} V_{MN}V^{MN},\label{54}
%\end{align} 
The action containing the first order of the $h^{\mu\nu}$ and the second order of $V_{\mu}$ is written as:
\begin{align}
    S=\dfrac{1}{2g^2_5}\int \frac{dz}{z}d^4x h^{\mu\alpha} \Big[\partial_\mu V^{(1)}_\nu \partial_\alpha V^{(2)\nu}-\partial_\mu V^{(1)}_\nu \partial^\nu V^{(2)}_\alpha-\partial_\nu V^{(1)}_\mu \partial_\alpha V^{(2)\nu}+\partial_\nu V^{(1)}_\mu \partial^\nu V^{(2)}_\alpha-\partial_z V^{(1)}_\mu \partial_z V^{(2)}_\alpha\Big].\label{eq: 5D-action-f2-V-V}
\end{align}
Using KK decomposition and Fourier transformation, we can write the action in momentum space:
\begin{align}
    S=-\dfrac{1}{2g^2_5}\int \frac{dz}{z}  d^4p_1 d^4p_2  \Big\{& h(q,z) V^{(1)}(p_1,z) V^{(2)}(p_2,z) h^{\mu\alpha} (q) \Big[p_{1\nu} p_{2}^\nu V^{(1)}_\mu(p_1)V^{(2)}_\alpha (p_2)-p_{1\mu} p_2^\nu V^{(1)}_\nu(p_1)V^{(2)}_\alpha (p_2)\nonumber
    \\
   &-p_{1\nu} p_{2\alpha} V^{(1)}_\mu (p_1) V^{(2)\nu} (p_2) +p_{1\mu} p_{2\alpha} V^{(1)}_\nu (p_1) V^{(2)\nu}(p_2)\Big] \nonumber
    \\
   &-h(q,z)\partial_z V^{(1)}(p_1,z) \partial_z V^{(2)}(p_2,z) h^{\mu\alpha}(q)V^{(1)}_\mu(p_1) V^{(2)}_\alpha(p_2)  \Big\},\label{eq:5D-f2-rho-rho-action}
\end{align}
where $q=-(p_1+p_2)$ due to $\delta(q+(p_1+p_2))$-function arising on $x$ integration. Here $p_1,p_2$ are the four-momenta of the $\rho^{(1)}, \rho^{(2)}$ mesons.  As is known from earlier works  on tensor meson couplings  \cite{Levy:1976, Renner:1971sj,  Mathieu:2020zpm}, the tensor-vector vector meson interaction amplitude contains five independent couplings due to the tensor structure of the tensor-meson current \cite{Levy:1976}:
\begin{align}
\mathcal{A}_{TV_1V_2}=\frac{g_{f_2V_1V_2}}{m_{f_2}^2}\tau^{\mu\nu}\Big\{\beta_0 \epsilon^{(1)}_{\mu}\epsilon^{(2)}_{\nu} +\beta_1 (\epsilon^{(1)}\cdot p_2)\epsilon^{(2)}_{\mu}p_{1\nu} +\beta_2 (\epsilon^{(2)}\cdot p_1)\epsilon^{(1)}_{\mu}p_{2\nu} +\beta_3 (\epsilon^{(1)}\cdot \epsilon^{(2})p_{1\mu}p_{2\nu} +\nonumber \\
+\delta(\epsilon^{(1)}\cdot p_2) (\epsilon^{(2)}\cdot p_1)  p_{1\mu}p_{2\nu}\Big\}\,,
\end{align}
with the polarization tensor $\tau^{\mu\nu}$ for spin-2 field and $\epsilon_{\mu}^{(1,2)}$-polarization vectors for massive vector fields. 

The first scenario,  the so-called  TMD model in which $\beta_0=p_1\cdot p_2$, $\beta_1= \beta_2 =- \beta_3 = -1$, and $\delta=0$ leads to the following Lagrangian
\begin{align} \label{eq:lag-f2rhorho}  L_{f_2\rho\rho}^{\text{TMD}}=g_{f_2\rho\rho}^{(1)}h^{\mu\alpha}V_{\mu\nu}V_{\alpha}^{\nu}=g_{f_2\rho\rho}^{(1)}h^{\mu\alpha}\Big(\partial_{\mu}V^{(1)}_{\nu}\partial_{\alpha}V^{(2)\nu}-\partial_{\nu}V^{(1)}_{\mu}\partial_{\alpha}V^{(2)\nu}- \partial_{\mu}V^{(1)}_{\nu}\partial^{\nu}V^{(2)}_{\alpha}+\partial_{\nu}V^{(1)}_{\mu}\partial^{\nu}V^{(2)}_{\alpha}\Big)\,,
\end{align}
where the coupling $g_{f_2\rho\rho}^{(1)}$ is evaluated as follows
\begin{align}
   |g_{f_2\rho\rho}^{(1){HW}}|= \frac{4\pi m_{f_2}^2}{2g_5^2\sqrt{5}}\int_0^{z_m}  \dfrac{dz}{z}\Big[ h^{HW}(z)  \Big(v^{HW}(z)\Big)^2 
    \Big]=0.31\, \text{GeV}\,,\qquad 
    \\
 |g_{f_2\rho\rho}^{(1){SW}}|= \frac{4\pi m_{f_2}^2}{2g_5^2\sqrt{5}}\int_0^{\infty}  \dfrac{dz}{z}e^{-k^2z^2}\Big[ h^{SW}(z)  \Big(v^{SW}(z)\Big)^2 
    \Big]= 0.68\, \text{GeV}\,.
\end{align}
Here the factor $\frac{4\pi}{\sqrt{5}}$ is inserted to do a compatible matching with 4-dimensional theory \cite{Katz:2005ir}.

In the case of the so-called minimal model \cite{Mathieu:2020zpm}, where all momentum dependent terms are neglected  $\Big(\beta_0=m_{f_2}^2, \beta_1=\beta_2=\beta_3=\delta=0\Big)$, the 4D Lagrangian has a simple form 
\begin{align}\label{eq:f2rhorho}
L_{f_2\rho\rho}^{\text{min}}=g_{f_2\rho\rho}^{(0)}h^{\mu\alpha}V_{\mu}V_{\alpha}\,.
\end{align}
Numerical estimation of the minimal coupling constant within the HW  model reads from the last term of Eq. \eqref{eq:5D-f2-rho-rho-action} :
\begin{align}
    |g_{f_2\rho\rho}^{(0) {HW}}|=\frac{4\pi}{2g_5^2\sqrt{5}}\int_0^{z_m}  \dfrac{dz}{z}\Big[ h^{HW}(z)  \Big(\partial_z v^{HW}(z)\Big)^2 
    \Big]=0.04\, \text{GeV}\,,
    \end{align}
Similarly, within the SW model:
    \begin{align}
 |g_{f_2\rho\rho}^{(0){SW}}|=\frac{4\pi}{2g_5^2\sqrt{5}}\int_0^{\infty}  \dfrac{dz}{z}e^{-k^2z^2}\Big[ h^{SW}(z)  \Big(\partial_z v^{SW}(z)\Big)^2 
    \Big]=0.08\,\text{GeV}\,.
\end{align}
%This value can be compared to the result of the chiral model \cite{Jafarzade:2022uqo} in which $g_{f_2\rho\rho}^{(0)}\approx 14 \, \text{GeV}$. Note that, the authors used the PDG data for the two pseudoscalar meson decays of the tensor meson to obtain the coupling of the chiral lagrangian which contains vector meson interactions as well, although it is not kinematically allowed.
Note that the Lagrangian describing the minimal model \eqref{eq:f2rhorho} can be extended to the following 5D form:
\begin{align}
L^{\text{min}-5D}_{f_2\rho\rho}=g_{f_2\rho\rho}^{(0)\prime}h^{MN}V_{M}V_N\,.
\end{align}
Numerical results for the coupling of the $g_{f_2\rho\rho}^{(0)\prime}$ in the HW and SW models are $g_{f_2\rho\rho}^{(0)\prime\,HW}=0.38\,\text{GeV}$ and $g_{f_2\rho\rho}^{(0)\prime\,SW}=0.84\,\text{GeV}$, respectively.

We move on to the next action describing the decay of the tensor meson into a vector meson and a photon. In this case, the action consists of the convolution of two field stress tensors $V_{MN}$ and $F_{MN}=\partial_M A_N-\partial_N A_M$ corresponding to vector meson and photon fields' stress tensors in the boundary \footnote{4D analogous of the action Eq. \eqref{eq:photon-vector-action} can be obtained via VMD \cite{Sakurai:1960ju,OConnell:1995nse}
by applying the following shift in the Lagrangian \eqref{eq:lag-f2rhorho}
\begin{equation}
V_{\mu\nu}\rightarrow V_{\mu\nu}+\frac{e}{g_{\rho}}\begin{pmatrix}
\frac{2}{3} & 0 & 0\\
0 & -\frac{1}{3} & 0\\
0 & 0 & -\frac{1}{3}
\end{pmatrix}F_{\mu\nu}\text{
.}\label{shiftvmd}%
\end{equation}
The electromagnetic field tensor is represented by $F_{\mu\nu}$. The electric coupling constant is denoted by $e$ and is equal to $\sqrt{4\pi\alpha}$. The photon-vector-meson transition is parameterized by $g_{\rho}$, which is approximately equal to 5.5. The holographic description of VMD is studied in \cite{Son:2003et}.}:
\begin{align}\label{eq:photon-vector-action}
    S=\int d^5 x \sqrt{g} F_{MN}V^{MN}. 
\end{align}

The KK  decomposition for the $A_M$ field will contain the photon profile function \cite{Katz:2005ir} 
\begin{align}\label{eq:photon-profile}
    A(p,z):=\gamma (z)=\frac{e}{\sqrt{3\pi^2}}\,.
\end{align}
In the first order of the $h^{\mu\nu}$, Eq. \eqref{eq:photon-vector-action} contains the tensor-vector-photon interaction terms similar to tensor-vector-vector coupling case and has the following explicit form:
\begin{align}\label{eq:coup-frhogam}
    S&=\int \frac{dz}{z} d^4p d^4p'  h(q,z) A(p,z) V(p',z) h^{\mu\alpha} (q) \times \nonumber
    \\
    &
    \times\Big[p_\nu p'^\nu A_\mu(p)V_\alpha (p')-p_\mu p'^\nu A_\nu(p)V_\alpha (p')-p_\nu p'_\alpha A_\mu (p) V^\nu (p')
   +p_\mu p'_\alpha A_\nu (p) V^\nu(p')\Big] \,.
\end{align}
%Note that the interactions containing the photon do not have minimal coupling within the holographic description because of the assumed photon profile function in Eq. \eqref{eq:photon-profile}.

%Therefore, we have automatically  eliminated the term that breaks gauge invariance- the minimal lagrangian for the photon.  
 The 5D action,  which describes tensor meson coupling to two photons in the 4D boundary, is written in the same way:
 \begin{align}\label{eq:coup-fgamgam}
    S&=\int d^5 x \sqrt{g} F_{MN}F^{MN}=\int \frac{dz}{z} d^4p d^4p' h(q,z) A(p,z) A(p',z) h^{\mu\alpha} (q) \times \nonumber
    \\
    &
    \times\Big[p_\nu p'^\nu A_\mu(p)A_\alpha (p')-p_\mu p'^\nu A_\nu(p)A_\alpha (p')-p_\nu p'_\alpha A_\mu (p) A^\nu (p')
   +p_\mu p'_\alpha A_\nu (p) A^\nu(p')\Big]\,. 
\end{align}
In an analogous way to the $g_{f_2\rho\rho}^{(1)}$ coupling, from the actions \eqref{eq:coup-frhogam} and \eqref{eq:coup-fgamgam} we write the integral representations for the $g_{f_2\gamma\rho}^{(1)}$ and $g_{f_2\gamma\gamma}^{(1)}$
\begin{align}
     g_{f_2\gamma\rho}^{(1)HW}= \frac{4\pi m_{f_2}^2}{\sqrt{5}}\int  \dfrac{dz}{z}\Big[ h^{HW}(z) \gamma(z)  v^{HW}(z) %+ \dfrac{1}{z^2} (\partial ^z V(z))^2 
    \Big]=1.16\, \text{GeV}\,, 
\end{align}
while the holographic model result for the two-photon coupling of the tensor meson \cite{Katz:2005ir} reads
\begin{align}
    g_{f_2\gamma\gamma}^{(1)HW}= \frac{4\pi m_{f_2}^2}{\sqrt{5}}\int  \dfrac{dz}{z}\Big[ h^{HW}(z) \Big(\gamma(z)\Big)^2  %+ \dfrac{1}{z^2} (\partial ^z V(z))^2 
    \Big]=0.06\, \text{GeV}\,.
\end{align}

\begin{table}[h]
\begin{tabular}{|c|c|c|c|c|}
\hline\renewcommand{\arraystretch}{2.}
  \centering
\text{Coupling} & $\Gamma_{f_2(1270)\rightarrow \pi^{+}\pi^{-}2\pi^{0}}$\cite{Mathieu:2020zpm}  & $\Gamma_{f_2\rightarrow \gamma\gamma}$\cite{Mathieu:2020zpm} &\text{Chiral model}\cite{Jafarzade:2022uqo} & \text{Our Result (HW)}  \\ \hline
%$g_{f_2\rho\rho}$ & $54.15$  & $10.96$ & $26.65$ & $19.40$ \\ \hline
$g_{f_2\gamma\rho}^{(1)}/m_{f_2}$ & $3.32$ & $0.68$ & $2.93$ &  $0.90$ \\ \hline
\end{tabular}
\caption{The HW model estimation for the coupling vs results derived from Refs. \cite{Mathieu:2020zpm,Jafarzade:2022uqo}}
\label{tab:gamrho}
\end{table}

Considering the following experimental results given in PDG \cite{Workman:2022ynf}
\begin{align*}   \Gamma_{f_2(1270)\longrightarrow \rho^0\rho^0+2\rho^-\rho^+}\simeq  \Gamma_{f_2(1270)\longrightarrow \pi^{+}\pi^{-}2\pi^{0}} \approx 19.5 \,\text{MeV}\,,\qquad\Gamma_{f_2\rightarrow \gamma\gamma}=2.6\pm 0.5\, \text{keV,}
\end{align*}
authors in Ref.\cite{Mathieu:2020zpm} extract the coupling $g_{f_2\gamma\rho}$ in Table \ref{tab:gamrho} using the following relation derived under the  Vector Meson Dominance (VMD) assumption:
%\begin{align}
 %    \text{TMD}:\quad g^{1}_{f_2\rho\gamma}=3.32\,,\quad g_{f_2\gamma\rho}^{2}=0.684
%\end{align}
%which imply
%\begin{align}
 %   \text{TMD}:\qquad g_{f_2\rho\rho}^{1}=54.15\,,\quad g_{f_2\rho\rho}^{2}=10.96
%\end{align}
\begin{align}
 g_{f_2\rho\rho}=\frac{m_\rho}{g_\rho}\frac{g_{f_2\gamma\rho}}{\sqrt{4\pi\alpha}}\,.
\end{align}
The following relation for the couplings of the radiative decays is valid within the quark model
    \begin{align}    g_{f_2\gamma\rho}=\frac{g_{f_2\gamma\gamma}}{4\pi\alpha}\frac{1}{\Big(\frac{g_\rho}{m_\rho}+\frac{1}{3}\frac{g_\omega}{m_\omega}\Big)}
\end{align}
the decay constants for the vector fields $\rho(770)$ and $\omega(782)$ are obtained from the experiment
\begin{align}
  g_{V}= \sqrt{ \frac{3m_V\Gamma_{V\rightarrow e^{-}+e^{+}}}{4\pi\alpha^2}}
\end{align}
Decay rate $\Gamma_{f_2\rightarrow\gamma\gamma}=2.6$ keV implies
\begin{align}    g_{f_{2}\gamma\gamma}=\sqrt{\frac{320 \pi \Gamma_{f_2\rightarrow\gamma\gamma} }{m_{f2}}}=0.045
\end{align}
which coincides with the HW AdS/QCD model in Ref. \cite{Katz:2005ir}. Considering the relation between the radiative decays of the tensor meson in equation A.46 of Ref. \cite{Oh:2003aw}, we obtain $\Gamma_{f_2\rightarrow\rho\gamma}=0.6\,\text{MeV}$ which is compatible with the prediction of the chiral model $0.7\pm 1.7\,\text{MeV}$ in Ref. \cite{Jafarzade:2022uqo}.

\section{Conclusion and future directions}\label{concluison}
We studied the holographic description of the tensor meson by exploring its hadronic and radiative coupling constants. The results are consistent with those obtained from Dispersion Relations and amplitude methods, and the holographic approach still remains effective in describing the couplings of tensor mesons. Our results show that the coupling for tensor meson-nucleon interaction coupling with one derivative term is the dominant one compared to the two-derivative case. The formalism presented here can be extended to describe the tensor glueball, which could be tested in the future Panda experiment \cite{PANDA:2009yku}. Motivated by the experimental observation of the 4-pion decay of $f_2(1270)$ (the second dominant decay channel after the two-pion channel), we have computed the coupling ($g^{(i)}_{f_2\rho\rho}$) within the holographic model. Radiative decay channels such as $f_2(1270)\rightarrow\rho(770)\gamma$, which is obtained via the VMD of the tensor-vector interaction Lagrangian, can be interesting for future experiment measurements, such as the GlueX experiment \cite{GlueX:2015eeb} at Jefferson Lab, to effectively test our prediction for the coupling. We predict the decay width for this channel to be around $0.6\, \text{MeV}$, which is consistent with the prediction presented in  Ref. \cite{Jafarzade:2022uqo}. One can study the finite temperature effects \cite{Mamedov:2021dpv,Mamedov:2021hkb}, the tensor form factors of nucleons given in Ref. \cite{Hagler:2009ni,Aliev:2011ku,Erkol:2011iw} and the tensor meson decay constant  \cite{MartinContreras:2019kah} within our model. 

\section*{Acknowledgements}
The authors thank Tahmasib Aliev for the useful discussion. S.J. thanks Francesco Giacosa for earlier collaborations on tensor mesons and Astrid Hiller Blin for fruitful discussion. S.J.  acknowledges the financial support through the project ``Development Accelerator of the Jan Kochanowski University of Kielce'', co-financed by the European Union under the European Social Fund, with no. POWR.03.05. 00-00-Z212 / 18 and partial support by the Polish National Science Centre (NCN) through the OPUS
project 2019/33/B/ST2/00613.

\appendix
\section{The SW tensor meson profile function revisited}\label{App-A}
In the SW model, the gravity action in the second-order perturbation $h_{\mu\nu}$ contains an additional exponential warping factor \cite{Abidin:2009hr}
\begin{align}
    S_{h}^{SW}=-\int d^5x\dfrac{e^{-2\kappa_1^2z^2}}{4z^3}(h_{\mu\nu,z}h^{\mu\nu}_{,z}+h_{\mu\nu}\Box h^{\mu\nu}),\label{25}
\end{align}
where the transverse-traceless gauge conditions $\partial^\mu h_{\mu\nu}=0$, and $h^\mu_\mu =0$ have been imposed. Solution for the perturbation can be found in the form $h_{\mu\nu}(x,z) = h_{\mu\nu}(x)h(x,z)$, and the linearized Einstein equation for the Fourier-transformed profile function $h(p,z)$ takes the form:  
\begin{align}
    \Bigg[\partial_z\big(\dfrac{e^{-2\kappa_1^2z^2}}{z^3}\partial_z\big)+\dfrac{e^{-2\kappa_1^2z^2}}{z^3}p^2\Bigg]h(p,z)=0. \label{26}
\end{align}

We get the solution to this equation expressed in terms of the hypergeometric function:
\begin{align}
    h(z)=N_1\big(\kappa_1z\big)^4 {}_{1}F_1\Big[2-\dfrac{m^2}{8\kappa_1^2};3;2\kappa_1^2z^2\Big]+N_2 G_{1,2}^{2,0}\left(
\begin{array}{c}
-\frac{m}{8\kappa_1^2}+1\\
0,2\\
\end{array}\middle\vert
-2\kappa_1^2z^2
\right),\label{27}
\end{align}
where $p^2=m^2$ has been taken which implies $m=4\kappa_1$ for the mass of spin-2 tensor meson in the SW model. The second term of the solution is non-normalizable, so, $N_2$ should vanish. To relate this solution with one in Eq. \eqref{eq:eom:sw-ten}, it is preferred to express the profile function in terms of the generalized Laguerre polynomials by taking into account the following relation between the hypergeometric function and Laguerre polynomial:
\begin{align}
{}_{1}F_1(-n,\alpha+1,x) = \frac{\alpha! n!}{(n+\alpha)!} L_n^\alpha(x).\label{28}
   \end{align}
For our solution $\alpha=2$ and $n=-\dfrac{m^2}{2-8\kappa_1^2}$. By taking into account these values the solution (\ref{27}) will be written in terms of Laguerre polynomials
   \begin{align}
 h_n^{SW}(z)=N\big(\kappa_1z\big)^4 \frac{2n!}{(n+2)!} L_n^2(2k_1^2z^2). \label{29}   
\end{align}        
The given solution can be interpreted as the $n$-th KK mode, and the mass spectrum of the modes can be found from $m^2_n=8\kappa_1^2(2+n)$. We observe that the profile function in (\ref{29}) coincides with one in \eqref{eq:sw-ten-prof-fun} with redefinition $2\kappa_1^2=\kappa^2$.

\section{The flavor invariant form of the Lagrangian}\label{App-B}
This appendix shows the flavor invariant form of the Lagrangian studied in Sec.\ref{sec:tens-vect coupl}. In the case of $N_f=2$, the
adjoint representation of $U(2)_f$ can be written as
\begin{align}
    \overline{2}\otimes 2 = 3 \oplus 1\,.
\end{align}
Considering the vector mesons as quark-anti-quark objects, we have triplets $\Big(\rho_{\mu}^{0},\rho_{\mu}^{+},\rho_{\mu}^{-}\Big)$ and the singlet $\Big(\omega_{\mu}\Big)$ within the following nonet $V_{\mu}$:
\begin{align}    V_{\mu\,i}^{j}=\overline{q}^{j}\gamma_{\mu}q_i=\begin{pmatrix}
    \overline{u}\gamma_{\mu} u &  \overline{u}\gamma_{\mu} d\\
     \overline{d}\gamma_{\mu} u &  \overline{d}\gamma_{\mu} d 
\end{pmatrix} \equiv \frac{1}{\sqrt{2}}\begin{pmatrix}
         \rho^{0}+\omega & \sqrt{2}\rho^{-}\\
         \sqrt{2}\rho^{+}&  -\rho^{0}+\omega
     \end{pmatrix}_{\mu}=\vec{\rho}_{\mu}\vec{\tau}+\omega_{\mu}\mathbb{1}_{2\times2}\,,
\end{align}
where $\vec{\tau}$'s are the three Pauli matrices. We have used the following substitutions:
\begin{align}
    \rho_{\mu}^0=\frac{1}{\sqrt{2}}\Big(\overline{u}\gamma_{\mu}u-\overline{d}\gamma_{\mu}d\Big)\,,\quad \rho_{\mu}^{+}=-i\,\overline{d}\gamma_{\mu}u \,,\quad \rho_{\mu}^{-}=i\,\overline{u}\gamma_{\mu}d\,,\quad \omega_{\mu}^0=\frac{1}{\sqrt{2}}\Big(\overline{u}\gamma_{\mu}u+\overline{d}\gamma_{\mu}d\Big)\,.
\end{align}
Thus, we can write $V_{\mu}:=V_{\mu}^at^a$ where $V_{\mu}^a=\{\Vec{\rho},\omega\}$ and $t^a=\{\Vec{\tau},\mathbb{1}_{2\times2}\}$. A matrix containing the vector fields is invariant under the flavor symmetry $U(2)_f$ such that
\begin{align}
    V_{\mu}\rightarrow U V_{\mu} U^\dagger\,,
\end{align}
where $U$'s are the elements of the unitary group $U(2)_f$ satisfy
the condition
\begin{align}    U^{\dagger}U=\mathbb{1}_{2\times2}\,.
\end{align}
Similarly, for the tensor fields, construct the flavor invariant matrix $h_{\mu\nu}$
\begin{align}    h_{\mu\nu}=\frac{1}{\sqrt{2}}\begin{pmatrix}
         a_2^{0}+f_2 & \sqrt{2}a_2^{-}\\
         \sqrt{2}a_2^{+}&  -a_2^{0}+f_2     \end{pmatrix}_{\mu\nu}=\vec{a}_{2\,\mu\nu}\vec{\tau}+f_{2\,\mu}\mathbb{1}_{2\times2}
\end{align}
which is invariant under $U(2)_f$
\begin{align}
    h_{\mu\nu}\rightarrow U h_{\mu\nu} U^\dagger\,.
\end{align}
The following lagrangian describes the interaction between the tensor and vector meson and is invariant under flavor symmetry $U(2)_f$
\begin{align}    \mathcal{L}&=g^{(0)}\text{Tr}\Big[h^{\mu\nu}V_{\mu}V_{\nu}\Big]+g^{(1)}\text{Tr}\Big[h^{\mu\nu}V_{\mu\alpha}V^{\alpha}_{\nu}\Big]=\\\nonumber
&=\frac{g_{f_2\rho\rho}^{(0)}}{\sqrt{2}}f_2^{\mu\nu}\Big(\rho_{\mu}^0\rho_{\nu}^0+2\rho_{\mu}^{+}\rho_{\nu}^{-}\Big)+\frac{g_{f_2\rho\rho}^{(1)}}{\sqrt{2}}f_2^{\mu\nu}\Big(\rho_{\mu\alpha}^0\rho_{\nu}^{0\,\alpha}+2\rho_{\mu\alpha}^{+}\rho_{\nu}^{-\,\alpha}\Big)+\cdots \,.
\end{align}

\bibliography{ads-qcd}
\end{document}